\documentclass[fleqn,usenatbib]{mnras}

\usepackage{newtxtext,newtxmath}
\usepackage[T1]{fontenc}

\DeclareRobustCommand{\VAN}[3]{#2}
\let\VANthebibliography\thebibliography
\def\thebibliography{\DeclareRobustCommand{\VAN}[3]{##3}\VANthebibliography}

\usepackage{graphicx}	% Including figure files
\usepackage{amsmath}	% Advanced maths commands
\def\dif{\mathrm{d}}
\def\xx{\boldsymbol{x}}
\def\kk{\boldsymbol{k}}
\def\qq{\boldsymbol{q}}

\def\pp{\boldsymbol{p}}
\def\PPsi{\boldsymbol{\Psi}}

\newcommand{\nenya}{{\tt Nenya }}
\newcommand{\vilya}{{\tt Vilya }}
\newcommand{\narya}{{\tt Narya }}
\newcommand{\theone}{{\tt TheOne }}

\newcommand{\paf}{``Paired-\&-Fixed''}

\newcommand{\Msun}{ h^{-1}{\rm M_{ \odot}}}
\newcommand{\hMpc}{ h^{-1}{\rm Mpc}}
\newcommand{\hkpc}{ h^{-1}{\rm kpc}}
\newcommand{\ihMpc}{ h\,{\rm Mpc}^{-1}}

\newcommand{\ensavg}[1]{\left\langle #1 \right\rangle}

\usepackage{etoolbox}
\title[Emulating biased tracers]{The BACCO simulation project: biased tracers in real space}

\author[Zennaro et al.]{
Matteo Zennaro,$^{1}$\thanks{E-mail:matteo\_zennaro001@ehu.eus}
Raul E. Angulo,$^{1,2}$\thanks{E-mail:reangulo@dipc.org}
Marcos Pellejero-Ib\'a\~nez,$^{1}$
Jens St\"ucker,$^{1}$
\newauthor
Sergio Contreras,$^{1}$
and Giovanni Aric\`o$^{1,3}$
\\
$^{1}$Donostia International Physics Center (DIPC), Paseo Manuel de Lardizabal, 4, 20018, Donostia-San Sebasti\'an, Guipuzkoa, Spain.\\
$^{2}$IKERBASQUE, Basque Foundation for Science, 48013, Bilbao, Spain.\\
$^{3}$Universidad de Zaragoza, Pedro Cerbuna 12, 50009 Zaragoza, Spain.
}

\date{Accepted XXX. Received YYY; in original form ZZZ}

\pubyear{2020}

\begin{document}
\label{firstpage}
\pagerange{\pageref{firstpage}--\pageref{lastpage}}
\maketitle

% Abstract of the paper (250 words max)
\begin{abstract}
We present an emulator for the two-point clustering of biased tracers in real space. We construct this emulator using neural networks calibrated with more than $400$ cosmological models in a 8-dimensional cosmological parameter space that includes massive neutrinos an dynamical dark energy. The properties of biased tracers are described via a Lagrangian perturbative bias expansion which is advected to Eulerian space using the displacement field of numerical simulations. The cosmology-dependence is captured thanks to a cosmology-rescaling algorithm. We show that our emulator is capable of describing the power spectrum of galaxy formation simulations for a sample mimicking that of a typical Emission-Line survey at $z\sim1$ with an accuracy of $1-2\%$ up to nonlinear scales $k\sim0.7\ihMpc$.
\end{abstract}


%%%%%%%%%%%%%%%%%%%%%%%%%%%%%%%%%%%%%%%%%%%%%%%%%%

%%%%%%%%%%%%%%%%% BODY OF PAPER %%%%%%%%%%%%%%%%%%

\section{Introduction}
The observed spatial distribution of galaxies and quasars offers an extremely valuable window to the physics of the universe. For instance, their clustering as a function of scale depends on the early universe physics, where properties of the primordial fluctuation as well as the parameters of the cosmological model leave distinctive signatures. Similarly, the relation between cosmic velocities and densities, encoded in the so-called redshift space distortions, offers a pathway to constrain the nature of gravity and the growth of structure \cite[e.g.][]{Kaiser1987,GuzzoEta2008}. Finally, baryonic acoustic oscillations \cite[e.g.][]{1998ApJ...496..605E} can be employed as a standard ruler to measure the expansion history of the universe, offering an opportunity to better constrain the properties of dark energy \cite[see e.g.][]{2013PhRvD..88f3537D,2013MNRAS.432.1928T,2015MNRAS.454.4326P}.

There is a large number of ongoing observational campaigns that will take on the wealth of information encoded in clustering (e.g. Euclid, see the review by \citealt[][]{Euclid}, DESI, \citealt[][]{DESI, DESI_2}, and J-PAS \citealt[][]{BonoliEtal2020}). These observations will map the position and shapes of hundreds of millions of galaxies up to tens of thousands of square degrees \cite[e.g.][]{BARTELMANN2001291, Takada_2014, Ivezi_2019, Dore2019WFIRST, Mandelbaum_2018}. Although it will be crucial to keep statistical uncertainties and observational systematic errors under control, the actual limitation in exploiting the surveys will arise from the precision with which the observed distribution of the luminous tracers can be modelled.

The challenge of modelling galaxy clustering arises from the nonlinearity of the physics involved. Although the early seeds of structure formation can be predicted accurately and efficiently using linearised Bolztmann-Einstein equations \citep{LewisChallinorLasenby2000,Lesgourgues2011}, the subsequent gravitational evolution will create nonlinearities which will dominate on small scales. Furthermore, galaxies and quasars are expected to form in particular regions of the universe with an efficiency that depends on the details of the galaxy formation physics involved.

The most accurate way to follow all these processes is provided by numerical simulations \citep[see, e.g.][for a review]{KuhlenVogelsbergerAngulo2012}. In these, fluctuations in the early universe are represented by a set of $N$-body particles which are then evolved solving their equations of motion. Recent advances in the field have shown that different codes and numerical approaches converge in the predictions of the nonlinear matter power spectrum to better than 2\% down to scales of $k\sim 10\ihMpc$ \citep{SchneiderEtal2016,SpringelEtal2020,Garrison_2018,AnguloEtal2020}. On the other hand, simulations are computationally expensive, thus it is typically not possible to carry them out for more than a handful of different choices of cosmological parameters. Nevertheless, several approaches have been suggested in the literature to speed up their production, \citep[e.g.][]{2002MNRAS.331..587M, 2013JCAP...06..036T, 2016MNRAS.459.2327I}, or to interpolate among the outputs of simulations \citep{HeitmannEtal2014,LiuEtal2018,NishimichiEtal2019,DeRoseEtal2019,GiblinEtal2019,EuclidEmu2019,WibkingEtal2019,WintherEtal2019,AnguloEtal2020,EuclidEmu2020}.

Another problem that arises from a potential modelling of galaxy clustering from numerical simulation is related to galaxy formation. Although there has been great progress in understanding the formation of galaxies, correlations with halo properties and the impact of various processes, predictions for galaxy properties are still uncertain and model dependent. This has the drawback that, when compared to observations, small uncertainties in galaxy modelling could heavily bias cosmological constraints. An alternative could be marginalization over galaxy formation. Though this is in principle possible, it would add much higher computational demands, and a poor or incomplete modelling of galaxy properties could still lead to biased cosmological inferences.

Perturbation theory offers a very attractive alternative. By solving analytically the relevant fluid equations, predictions for the distribution of matter down to quasilinear scales can be obtained efficiently \citep[e.g.][]{BernardeauEtal2002} and with control of theoretical uncertainties \cite[see e.g.][]{ChudaykinEtal2020}. The relation between biased tracers and the underlying matter field can also be treated perturbatively \citep{McDonald_2009, 2018PhR...733....1D, Fujita_2020}. By including all possible dependences to a given order allowed by symmetries, it is possible to extend the predictions for any biased tracers, without making an explicit connection with particular galaxy formation physics. Recent advances in these fields have significantly improved the accuracy of these predictions, which can typically reach scales of $k \sim 0.2\ihMpc$ (\citealt{Baumann_2012}, \citealt{Baldauf_2016}, \citealt{Vlah_2016}, \citealt{Ivanov_2020}, \citealt{d_Amico_2020}, \citealt{Colas_2020}, \citealt{Nishimichi2020}, \citealt{ChenVlahWhite2020} ). Although this is a remarkable achievement, these predictions still fall short with respect to the accuracy expected from future galaxy surveys (\citealt{Blas_2014}, \citealt{McQuinn_2016}). This approach has been extensively used under the assumption of perturbed dynamics in cosmological parameter estimation (see e.g. \citealt{Chuang_2017} and \citealt{Pellejero-Ibanez_2017}).

%,  and has also been recently tested on $N$-body-like computed dynamics in \cite{ModiChenWhite2020} finding sub-percentage accuracy in the real space power and cross spectrum until scales of $k\approx 1 h{\rm{Mpc}^{-1}}$

In this paper we combine both approaches -- numerical simulations and perturbation theory -- to create a framework that inherits the accuracy of numerical simulations with the flexibility of a perturbative bias expansion, which can reach even small nonlinear scales while being agnostic to galaxy formation physics and details of any given particular observational surveys. As shown by \cite{ModiChenWhite2020}, this approach has the potential to accurately describe galaxy clustering down to smaller scales than those typically reached by perturbation theory.

We are able to do this by combining the benefits of several recent developments. On the simulation side, we employ large $N$-body simulations, with accurate force and mass resolution, which have significant noise suppression by using special initial conditions, and have been carefully designed to be used in combinations with cosmology-rescaling algorithms. These allow to densely cover a target parameter space with the equivalent of hundreds to thousands of simulations. We combine these with feed forward neural networks which allow us to quickly predict non-linear fields while varying any cosmological parameter -- including neutrinos and dynamical dark energy -- within defined regions in cosmological parameter space. On the perturbative side, we employ a Lagrangian bias expansion up to second order which captures dependences with the local density and tidal fields, and that includes a higher-order derivative bias parameters, capturing the non-locality of the galaxy formation process. Using the non-linear displacement field of the rescaled simulations the Lagrangian bias descriptions are advected to Eulerian space where the different fields can be combined to make predictions for the non-linear power spectrum of biased tracers.

This paper is structured as follows. In \S\ref{sec:numerical} we describe the variety of numerical tools we employ, including numerical simulations and the power spectrum calculation, and we validate the cosmology-rescaling approach. We also describe the implementation of the Lagrangian bias expansion. In \S\ref{sec:lpt} we describe a perturbative solution for the matter fields we consider, which we will employ to complement our numerical approach. In \S\ref{sec:emulator} we provide details of our emulation via neural networks, including its validation and estimation of its accuracy. In \S\ref{sec:application} we present an application of our emulator by fitting the power spectrum of galaxies mimicking a Star Formation Rate selected (SFR) sample at $z=1$. We conclude in \S\ref{sec:conclusions}.

\section{Numerical Methods}
\label{sec:numerical}

Here we present the main numerical methods underlying this work. Specifically, we describe our simulations in \S\ref{sec:simulations} and how we model biased tracers in \S\ref{sec:bias}. We finish by validating the cosmology-rescaling approach for our purposes in \S\ref{sec:rescaling}.

\subsection{Simulations}
\label{sec:simulations}

We will employ two sets of simulations. The first one, referred to as the BACCO simulations will be the core of our biased-tracers emulators. The second suite will be employed to test the accuracy of our predictions.

\begin{table}
  \centering
  \begin{tabular}{cc|cccccccc} % eight columns, alignment for each
     \hline
     Cosmology &  $\Omega_{\rm cdm}$ & $\Omega_{\rm b}$ & $h$ & $n_{\rm s}$\\
     \hline
    \nenya  & 0.265 & 0.050 & 0.60 & 1.01 \\
    \narya  & 0.310 & 0.050 & 0.70 & 1.01 \\
    \vilya  & 0.210 & 0.060 & 0.65 & 0.92 \\
    \theone & 0.259 & 0.048 & 0.68 & 0.96 \\
     \hline
     \end{tabular}
    \caption{Cosmological parameters of the four cosmologies simulated in the BACCO project. All the cosmologies assume a flat geometry, no massive neutrinos ($M_{\nu}=0$ eV), a dark energy equation of state with $w_0=-1$ and $w_a=0$, an amplitude of cold matter fluctuations $\sigma_8=0.9$, and optical depth at recombination $\tau=0.0952$.}
  \label{tab:parameters_table}
\end{table}

\subsubsection{BACCO Simulations}

Our main suite of simulations corresponds to the core of the BACCO simulation projects. These corresponds to 8 large $N$-body simulations specially designed to cover a wide range of cosmological parameters when combined with cosmology-rescaling. These simulations were presented in \cite{AnguloEtal2020}, we here simply provide a recap of their main characteristics.

Specifically, the BACCO simulations are a set of gravity-only simulations with a box of sidelength $L=1440\,\hMpc$ resolved with $4320^3$ particles, which implies a mass resolution of $m_p \sim 3 \times 10^9\,\Msun$. These simulations are carried out in pairs at 4 distinct sets of cosmological parameters. The parameter values, provided in Table \ref{tab:parameters_table}, were chosen so that, in combination, these simulations can efficiently cover a region of approximately $10\sigma$ around the best fit values obtained by the analysis of the Planck satellite. We refer to \cite{ContrerasEtal2020} for details on how these parameters were chosen. For each of these cosmologies we carried out two simulations whose initial Fourier amplitudes have been fixed but their initial phases are inverted. As shown by \cite{AnguloPontzen2016}, this configuration allows for a dramatic reduction of the noise in the power spectrum due to cosmic variance -- by more than two orders of magnitude for $k<0.1\ihMpc$.

The gravitational evolution was carried out with an updated version of {\tt L-Gadget3} \citep{Springel2005,AnguloEtal2020}, and employing a Plummer-equivalent softening length of $5\hkpc$. Numerical parameters were chosen so that the power spectrum displays a convergence better than 2\% at $k\sim 10\,\ihMpc$. In fact, as shown in \cite{AnguloEtal2020}, such configurations agree to better than 2\% with other state-of-the-art codes in a realization of the "Euclid Simulation Challenge" \citep{SchneiderEtal2016}.

\subsubsection{Test Suite of Simulations}

A key aspect of our framework is the ability of employing the 4 BACCO cosmologies to sample hundreds of different cosmologies using a cosmology-rescaling.

To test the accuracy of such rescaling, we will employ a suite of $35$ independent $N$-body simulations. Each of these simulations consists of $1536^3$ particles in a cubic volume of $512\hMpc$ a side. Naturally, these simulations feature a much smaller volume than our main BACCO simulations, however they have identical numerical parameters in terms of mass resolution and force resolution. As for our main suite, these were carried out with the latest version of {\tt L-Gadget3} and were initialised in pairs using the approach of \cite{AnguloPontzen2016} and employing 2LPT at $z=49$.

To further improve the accuracy of our comparison, we have carried out a version of the BACCO simulations but with a volume and initial phase field matching those of each of our test simulations. In this way, we reduce significantly the role of cosmic variance and allow for an accurate testing.

The cosmologies of our test suite vary systematically one of eight cosmological parameters $\boldsymbol{\vartheta} = \{\Omega_{\rm m}, \Omega_{\rm b}, \sigma_8, n_s, h, M_{\nu}, w_0, w_a \}$, over the same range in which we will build our emulator. When testing the accuracy of our predictions we will compare the results of rescaling these BACCO simulations against simulations carried out directly with the target cosmology.

\begin{figure*}
\includegraphics[width=\textwidth]{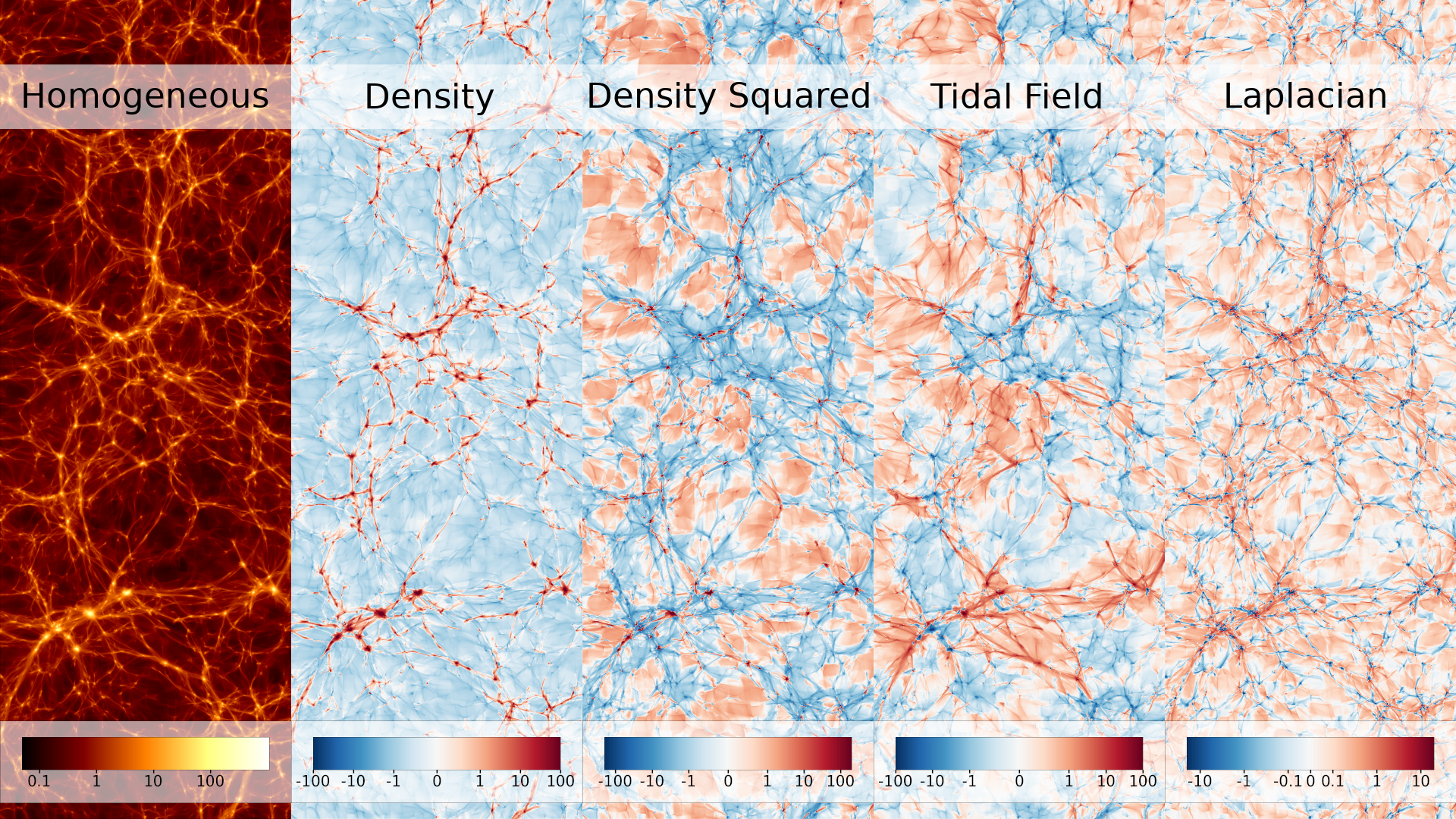}
    \caption{Visualization of the different Lagrangian fields that have been advected to Eulerian space. Each panel corresponds to a projection of the same $100 \times 281\times 25 \, \hMpc$ volume. The first panel corresponds to a uniform weighting -- thereby recovering the dark matter density field, whereas the other panels are weighted by different components of the Lagrangian linear density field. One can see how the different Lagrangian bias components emphasize different aspects of the large-scale structure -- for example the linear density weighting seems to emphasize filaments and clusters in the cosmic web and the density squared weighting emphasizes the regions associated with the most massive clusters (small red points).  \label{fig:fields}}
\end{figure*}

\begin{figure*}
\includegraphics[width=\textwidth]{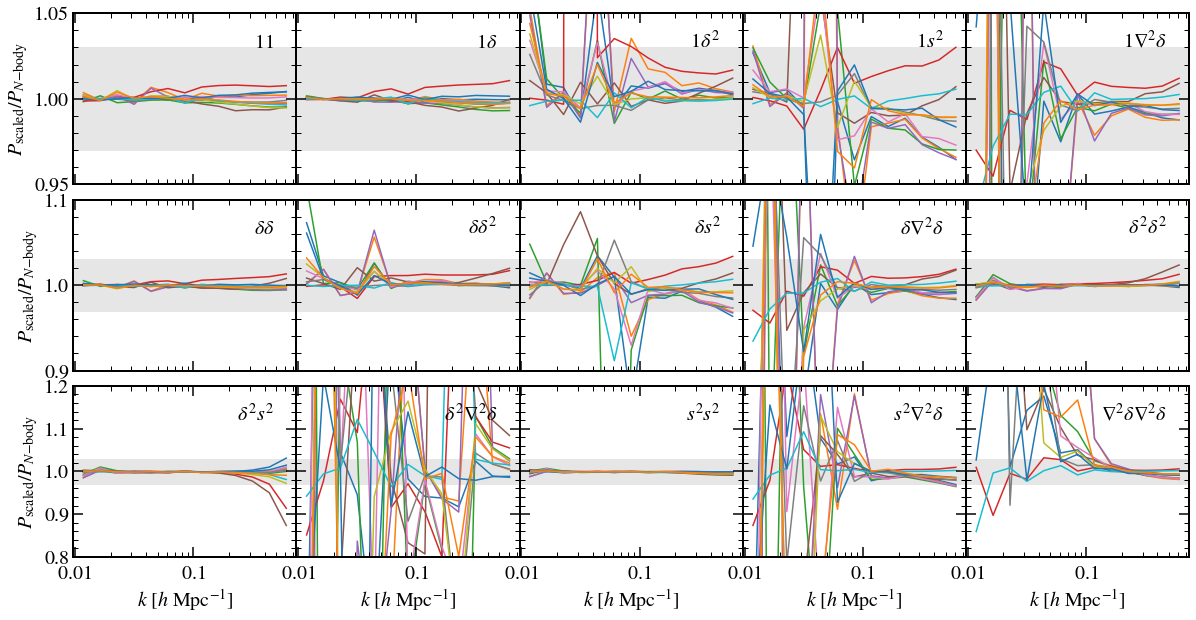}
\caption{Comparison showing the ratio of the power spectrum of linear Lagrangian fields advected to Eulerian coordinates as predicted by $N$-body simulation and by cosmology-rescaled simulations. Each panel displays the results for the cross-spectra of different linear fields, $P_{ij}$, as indicated by the legend; whereas  lines of different colours shows different cosmological models in our test suite, which includes dynamical dark energy an massive neutrinos.
\label{fig:rescaling}}
\end{figure*}

\subsection{Modelling biased tracers}
\label{sec:bias}

To model biased tracers, we will consider a Lagrangian bias expansion. We refer to \cite{DesjacquesJeongSchmidt2018} for a review of the perturbative bias formalism. As discussed in the introduction, the use of a general biasing formalism will allow us to model the clustering of any tracer of the underlying matter field, without making any strong assumptions about galaxy formation physics.

Specifically, we will describe the overdensity of objects in Lagrangian space, $\delta_{\rm g}(\qq)$, as a second-order expansion in the linear matter overdensity $\delta(\qq)$ and include potentially non-local dependences via a higher-order derivative $\nabla^2\delta(\qq)$. Hence, the Lagrangian density field describes the distribution of biased tracers once weighted by the function:

\begin{equation}
    \begin{split}
        F(\qq) &= 1 + b_1 \delta(\qq) + b_2 \left[\delta^2(\qq) - \ensavg{\delta^2}\right]\\
        &+ b_{s^2}\left[s^2(\qq) - \ensavg{s^2}\right] + b_{\nabla^2\delta} \nabla^2\delta(\qq),
    \end{split}
    \label{eq::lag-bias-exp}
\end{equation}

\noindent where $1$ is a homogeneous field; $s^2(\qq) = s_{ij}(\qq)s_{ij}(\qq)$ is the shear field, defined by the tidal tensor $s_{ij}(\qq) = \partial_i\partial_j \Phi(\qq) - \delta_{ij} \delta(\qq)$, with $\Phi(\qq)$ the local matter gravitational potential. Here, $b_1, b_2, b_{s^2},$ and $b_{\nabla^2\delta}$ are the Lagrangian bias parameters, which we assume to be scale-independent in Lagrangian space. Note that it is possible to include higher order bias terms and derivatives \cite[e.g.][]{FujitaEtal2020}, albeit they are expected to have a small contribution for low and intermediate mass haloes (\citealt{Abidi_2018}, \citealt{Lazeyras_2019}).

To describe the tracer overdensity field in Eulerian space, $\delta_{\rm g}(\xx)$, the coordinate system needs to be advected $\xx = \qq + \PPsi(\qq)$, where $\PPsi(\qq)$ is referred to as the displacement field. In terms of the advected fields $\delta_i(\xx)$, the power spectrum can then be expressed in Eulerian space as

\begin{equation}
P_{\rm gg} = \sum_{i,j \in \{1,\delta,\delta^2,s^2,\nabla^2\delta\}} b_i b_j\, \langle |\delta_i(\kk) \delta_j^*(\kk)|\rangle,  \\
\end{equation}

\noindent where $\delta_i(\kk)$ corresponds to the Fourier transform of the advected field $\delta_i(\xx)$, and it is completely determined by 15 cross-spectra, $P_{ij} \equiv \langle |\delta_i(\kk) \delta_j^*(\kk)|\rangle$. Usually, these spectra are computed perturbatively up to, at least, the same order as the bias expansion. This has the advantage that predictions can be computed quickly and accurately as a function of cosmology, but only on relatively large scales.

In this work, we follow a different strategy and directly compute the 15 cross-spectra relevant for the 2nd-order bias expansion using the results of numerical simulations. As shown by \cite{ModiChenWhite2020}, this approach improves significantly the reach of scales accessible to analytic bias expressions and, specifically, was able to accurately describe the clustering of mock HOD galaxies down to scales of $0.6\ihMpc$.

Operationally, to implement this approach we create the 5 relevant fields -- $1$, $\delta(\qq)$, $\delta^2(\qq)$, $s^2(\qq)$, and $\nabla^2\delta(\qq)$ -- in Lagrangian coordinates using a grid of $1080^3$ points, using Fourier amplitudes and phases matching those of our $N$-body simulations. We then apply a Gaussian smoothing of size $k_{\rm d}0.75\hMpc$, mimicking possible exclusion effects and the non-locality of structure formation. The choice of the smoothing scale is anoher parameter of the model, and it slighly affects the values of the inferred quadratic bias parameters ($b_2$ and $b_{s^2}$) and of the nonlocal bias parameter ($b_{\nabla^2\delta}$). Here, we have chosen the value of the smoothing scale, $k_{\rm d}$, so that it roughly coincides with the Lagrangian size of typical dark matter halos. For more details on the smoothing scale, we refer the reader to the companion paper \citep{ZennaroEtal2022}, where we show that assuming different values of $k_{\rm d}$ does not impact our ability to model the measured galaxy power spectra for scales $k<k_{\rm d}$. We then compute a non-linear displacement field on the same Lagrangian field by following simulation particles initially located at the same grid points.
Finally, we compute the corresponding advected fields by using a cloud-in-cell assignment in Eulerian space where the position is only given by the cosmology-dependent displacement field and where the weight is given by the value of the respective bias expansion field.

In Figure \ref{fig:fields} we show a slice through each of these 5 fields in Eulerian space: the `homogeneous' is the field weighted by a constant $1$, the `density' is weighted by $\delta(\qq)$, the `quadratic density' field is weighted by $\left[\delta^2(\qq) - \ensavg{\delta^2}\right]$, the `tidal field' by $\left[s^2(\qq) - \ensavg{s^2}\right]$, and the `Laplacian' by $\nabla^2\delta(\qq)$.

This visualization corresponds to a $25\hMpc$ projection from a $100 \times 281\hMpc$ section of one of our simulations. The colour coding emphasises how different weighting gives more relevance to different density environments. Specifically, the $\delta$ component amplifies collapsed regions (such as nodes in the cosmic web). The $\delta^2$ term also emphasizes collapsed regions but exhibits a steeper weighting within their inner regions. Additionally, due to its squared nature, it also upweights voids. The tidal term enhances the sheet-like and filamentary structure of the cosmic web. Finally, the Laplacian term mimics any effects that smooth out the density profile of collapsed regions, such as gas pressure perturbations \citep{DesjacquesJeongSchmidt2018}.

A key aspect of our work is that the displacement field connecting Lagrangian and Eulerian spaces can be easily and accurately evaluated for different cosmologies with a cosmology rescaling algorithm. We will explore this in the following subsection.

\begin{figure*}
\includegraphics[width=\textwidth]{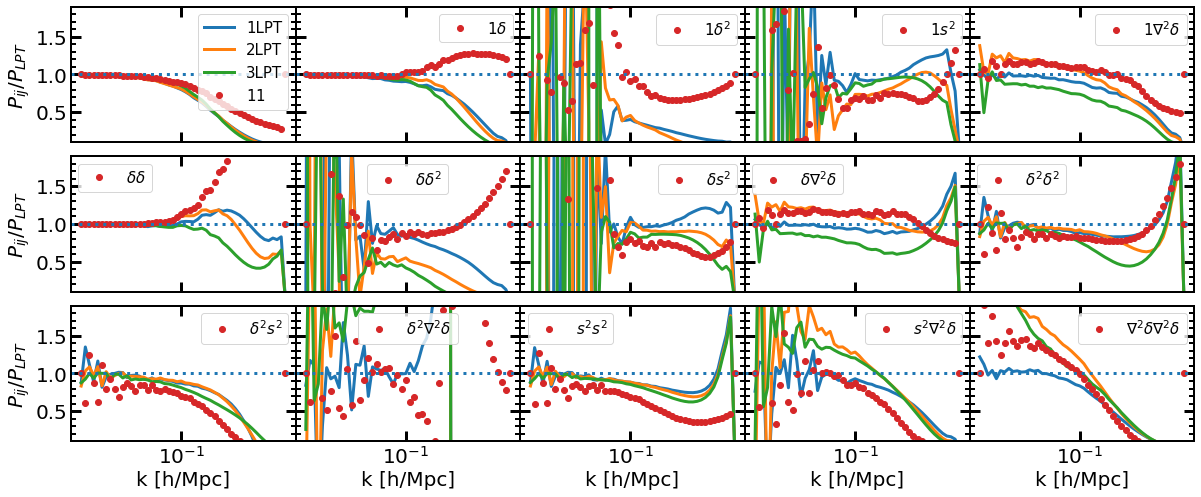}
\caption{ Comparison of the power spectra of linear Lagrangian field advected to Eulerian coordinates at $z=0$, as predicted by analytic LPT calculations, $P_{\rm LPT}$, and various kinds of simulations. Specifically, in each panel coloured lines show the results of using an ensemble of $1000$ realisations of $L=1500\hMpc$ advected using 1st, 2nd, or 3rd-order LPT displacements; whereas red symbols show the measurements from one of our $N$-body simulations. Each panel shows results for a different cross-spectra as indicated by the legend.
\label{fig:lpt-ratios}}
\end{figure*}

\subsection{Cosmology Rescaling}
\label{sec:rescaling}

To capture the cosmology dependence of the Lagrangian fields in Eulerian coordinates, we will employ the so-called cosmology-rescaling algorithms.

The main idea of such approach is that the nonlinear structure in a given cosmology can be mimicked by rescaling the outputs of a simulation carried out in a nearby cosmology. The rescaling is performed by transforming the length unit and by considering a different output time such that the linear variance of the field as a function of scale matches that of the rescaled simulation. Additionally, large-scale modes are modified by additionally displacing a given simulation particle with the difference of the 2LPT displacements in the original and target cosmologies. Velocities are modified in an analogous manner.

Cosmology-rescaling was originally proposed by \cite{AnguloWhite2010} and has been extensively tested in multiple subsequent studies. In particular, in \cite{ZennaroEtal2019} we extended the approach to massive neutrinos; in \cite{ContrerasEtal2020} we showed that the matter power spectrum of dark matter, haloes, and subhaloes can be retrieved to better than 3\% up to $k \sim 5\ihMpc$; and in \cite{OndaroEtal2022} that the halo mass function can be obtained to better than 2\% over the range $10^{12} - 10^{15}\, \Msun$. Additionally, this approach was employed by \cite{AnguloEtal2020} to construct an emulator for the nonlinear matter power spectrum, and by \cite{AricoEtal2020} to incorporate the effects of baryonic physics such as star formation, gas cooling and feedback.

Here, we will quantify the performance of cosmology-rescaling in predicting the cross-spectra of our Eulerian linear fields. For this, we will compare the measurements in our suite of test simulations (c.f. \S\ref{sec:simulations}) with those obtained by rescaling one of our BACCO simulations.

We present our results in Fig.~\ref{fig:rescaling} for the 15 possible combinations of our 5 Lagrangian fields at $z=0$, as indicated by the legend in each panel. Each of the lines displays the results for one of our test cosmologies. We recall that these cover a parameter space roughly set by a $10\sigma$ region around Planck's best fit values. Firstly, we can see that cosmology rescaling retrieves highly-accurate predictions, agreeing with those of direct $N$-body simulations to better than 1-2\% in most cases.
Although not shown here, we have checked that the precision of the method is even slightly better at higher redshifts.

The large-scale clustering is expected to be dominated by the terms involving the linear density and the homogeneous field -- $\langle11\rangle$, $\langle1\delta\rangle$, and $\langle\delta\delta\rangle$ -- these combinations are particularly well recovered, to better than 1\%, even on the smallest scales we consider.  On smaller scales, we expect fields involving $\delta^2$ -- the combinations $1\delta^2$, $\delta \delta^2$, and $\delta^2 \delta^2$ -- to also have relevant contributions. Similarly, for such fields, we also obtain highly accurate predictions, with deviations being typically below $\sim 2\%$ at $k\sim 1 \ihMpc$. Note that these fields have particularly low amplitudes on large scales, which is why the comparison becomes progressively noisier as we consider smaller wavenumbers.

Finally, we expect fields involving the tidal field -- $1s^2$, $\delta s^2$, $s^2 s^2$ -- to have subdominant contributions on all scales. Nevertheless, our predictions are still relatively accurate with typical discrepancies of less than 5\%. The least accurate predictions among all the cases where we can reliably perform the comparison is $\delta^2 s^2$, where the cosmology-rescaling can over- or under-predict the results of our test simulations by $5-10\%$.

We note that, even though these simulations have matching Fourier phases, some cross-spectra are particularly noisy on large scales. As we will see in the next sections, these are inherently noisy. These fields also display no benefit from the \paf\ method.

In summary, we have shown that, using a cosmology-rescaling, it is possible to accurately predict all the 15 cross-spectra involved in the perturbative bias expansion. This is remarkable as it opens up the possibility to compute predictions for a densely-sampled cosmological parameter space which should yield to an accurate emulation. Nevertheless, numerical predictions for some combination of fields are particularly noisy on large-scales (e.g. $\delta^2\nabla^2\delta$, which is zero at linear order even on large scales), which could pose a problem for a direct emulation. For this reason we will combine our numerical results with analytic calculations using Lagrangian perturbation theory, which will be the subject of our next section.

\section{Lagrangian Perturbation Theory}
\label{sec:lpt}

In the previous section, we have introduced the Lagrangian bias expansion, and in particular, the 15 cross-spectra of the different Lagrangian fields that enter the biased-tracers power spectrum. In this section, we model these 15 terms theoretically, with the aim of using these LPT predictions to reduce the noise present in the measurements from rescaled simulations, and provide us with a way of reducing the dynamical range of the quantities we want to emulate.

\subsection{LPT predictions}

To compare with quantities measured in simulations, we want each of the components of the field presented in Eq. (\ref{eq::lag-bias-exp}) to be evaluated at a given Eulerian position $\xx$. The field at that position will receive contributions from all Lagrangian positions that, after being displaced, end up in the same Eulerian position \citep[e.g.][]{Matsubara2008}. Therefore,

\begin{equation}
    1 + \delta_{F} (\xx) = \int \dif^3 \qq F(\qq) \delta_{\rm D}\left(\xx - \qq - \PPsi(\qq) \right),
    \label{eq::advection}
\end{equation}

\noindent where $F(\qq)$ can be any of the components of the field.

We obtain the overdensities created in Eulerian space from displacing a uniform distribution $F(\qq) = 1$, a linear density field $F(\qq) = \delta_{\rm L}(\qq)$, its square $F(\qq) = \delta^2_{\rm L}(\qq)$, the tidal field $F(\qq) = s^2(\qq)$ and the laplacian of the linear density field $F(\qq) = \nabla^2 \delta_{\rm L}(\qq)$. Note that we expand the displacement field $\PPsi \simeq \PPsi^{(1)} + \PPsi^{(2)} + \dots$, retaining only terms up to second order in power, so that in Fourier space
\begin{equation}
    \begin{split}
    \delta_{F}(\kk) = \int \dif^3 \qq e^{-i \kk \cdot \qq} F(\qq) &\lbrace 1 - i \kk \cdot \PPsi^{(1)}(\qq) - \dfrac{1}{2} [\kk \cdot \PPsi^{(1)}(\qq)]^2 \\
    &- i \kk \cdot \PPsi^{(2)} \rbrace,
    \end{split}
    \label{eq::advection-kspace}
\end{equation}
ignoring an additional Dirac's delta $\delta_{D}(\kk)$ that would only contribute to the wavemode $\kk=0$. Note that the first and second order displacement field can be easily connected to the linear overdensity in Fourier space, as
\begin{equation}
    \begin{split}
        \PPsi^{(1)} (\kk) &= -i \dfrac{\kk}{k^2} \delta(\kk).\\
        \PPsi^{(2)} (\kk) &= -\dfrac{i}{2}\int\dfrac{\dif^3 \kk_1}{(2\pi)^3} \boldsymbol{L}_2(\kk_1, \kk-\kk_1, \kk) \delta(\kk_1)\delta(\kk-\kk_1),
    \end{split}
\end{equation}
with
\begin{equation}
    \boldsymbol{L}_2(\kk_1, \kk_2, \kk) = \dfrac{3}{7}\dfrac{\kk}{k^2}\left[ 1 - \left( \dfrac{\kk_1\cdot\kk_2}{k_1 k_2}\right)^2 \right].
\end{equation}

After advecting these fields with Eq. (\ref{eq::advection-kspace}) we combine them into the respective cross spectra, retaining only terms of order $(11)$ and $(22)$, obtaining contributions of the form
\begin{equation}
    \begin{split}
        P_{ij}(k) &= A_{ij}(k) P(k)\\
        &+ B_{ij}(k) \int \dfrac{\dif^3 \pp}{(2 \pi)^3} P(\pp) P(|\kk - \pp|) C_{ij}(\pp, \kk),
    \end{split}
    \label{eq::pt-terms-general}
\end{equation}

\noindent where the prefactors $A_{ij}(k)$ and $B_{ij}(k)$, and the kernels $C_{ij}(\pp, \kk)$ are different for each combination of the fields and are reported in appendix \ref{sec::lpt-predictions}.

Numerically, we compute all LTP integrals using a doubly-adaptive quadrature integration, as implemented in the Gnu Scientific Library, whose accuracy we have tested against a multi-dimensional numerical integration. Each integral takes approximately $0.2$ seconds. In Appendix \ref{sec::velocileptors-fastpt} we compare our calculations against two publicly-available code {\tt FastPT} \citep{McEwenEtal2016,FangEtal2017} and {\tt velocileptors} \citep{ChenVlahWhite2020}.

\subsection{LPT validation}
To validate our LPT expressions and their numerical evaluation, we will compare them against a suite of realizations of an initially homogeneous distribution advected to Eulerian space using Lagrangian perturbation theory at different orders.

Specifically, we construct a suite of $1000$ realizations of the relevant Lagrangian fields on $384^3$ grids in $L=1500\hMpc$ boxes, and using a cosmology compatible with current observational constraints. These grids are then advected to $z=0$ using either first, second, or third-order Lagrangian perturbation theory, following the implementation of \cite{MichauxEtal2020}.

We then measure the resulting cross-spectra using FFTs with $384^3$ grid points.
We show our results in Fig.~\ref{fig:lpt-ratios} where we display the measured spectra over our LPT analytic calculations. For comparison, we also display the corresponding measurements from one of our main BACCO simulation as red symbols.

Firstly, we can see that for most 15 power spectra, our analytic calculations and the 1LPT simulations agree remarkably well on large-scale -- the ratio approaches to unity for low wavenumers. This is an important validation of our calculations. Furthermore, for terms involving, $1$ and $\delta$, on large scales, there is also agreement with higher-order version of LPT and with the results of our full $N$-body simulation. On smaller scales ($k > 0.1 \ihMpc$) LPT breaks down and is no longer able to reproduce the $N$-body results; since on these scales LPT is no loger in its perturbative regime, higher orders are not guaranteed to yield any improvements, and, in fact, we can see some cases where 3LPT at small scales performs worse than lower order cases. Note that the exact scale at which LPT breaks is not the same for different fields, because the weighting scheme can amplify nonlinear structures, causing shell crossing to happen on larger scales.

In other cases, there is a disagreement between 1 and 2LPT simulations with the later agreeing with the $N$-body results. This suggest that to predict accurately these quantities, higher-orders in the displacement need to be included.  Yet in other instances, mostly those involving $\delta^2$, the LPT simulations appear to converge to a different large-scale limit than the $N$-body result, which points to shell-crossing and nonlinearities on small scales being important to correctly predict the amplitude on large-scales. We leave further investigation of this for future work.

Nevertheless, the discrepancies on large scales appear in fields which are already highly subdominant on large scales. Thus we conclude that our analytic LPT calculations are sufficiently accurate to complement our numerical measurements on large scales.

\section{Emulating biased tracers}
\label{sec:emulator}

%=============================================
\begin{figure*}
\includegraphics[width=\textwidth]{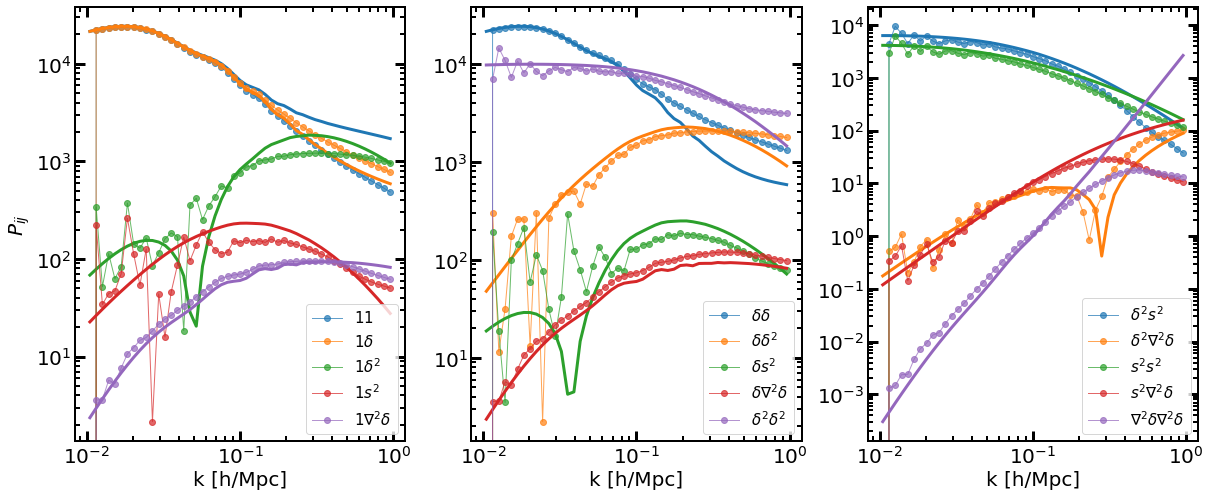}
\caption{ The cross spectra of various Lagrangian fields advected to Eulerian space, $P_{ij}$ where $i,j = \{1, \delta, \delta^2, s^2, \nabla^2 \delta\}$. Symbols display the measurements employing a randomly-selected cosmology from our training data at $z\sim 0$. We note that these measurements corresponds to cosmology-rescaled high-resolution simulations, which have been created with the \paf\ method. For comparison, solid lines show the predictions of our analytic LPT calculation for each respective cross-spectrum.
\label{fig:lpt}}
\end{figure*}
%=============================================

We now describe the procedure we follow to construct an emulator for real-space biased tracers. We first define our target cosmological parameter space in \S\ref{sec:params}, and then describe our core power spectrum data in \S\ref{sec:data}. We continue by discussing our neural network emulation strategy in \S\ref{sec:emulation} and finish by quantifying its accuracy in \S\ref{sec:validation}.

\subsection{Parameter Space}
\label{sec:params}

We will construct and validate our emulator over a hyper-volume in cosmological parameter space defined by the ranges:

\begin{eqnarray}
\label{eq:par_range}
\sigma_8                  &\in& [0.73, 0.9] \nonumber\\
\Omega_{\rm m}            &\in& [0.23, 0.4] \nonumber\\
\Omega_{\rm b}            &\in& [0.04, 0.06] \nonumber\\
n_{\rm s}                 &\in& [0.92, 1.01]\\
h  &                       \in& [0.6, 0.8] \nonumber\\
M_{\rm \nu}\,[{\rm eV}]   &\in& [0.0, 0.4] \nonumber\\
w_{0}                     &\in& [-1.15, -0.85] \nonumber\\
w_{\rm a}                 &\in& [-0.3, 0.3] \nonumber
\end{eqnarray}

\noindent where $\sigma_8$ is the {\it cold} mass linear mass variance in $8\,\hMpc$ spheres; $\Omega_{\rm m} $ and $\Omega_{\rm b}$ are the density of cold matter and baryons in units of the critical density of the Universe; $n_{\rm s}$ is the primordial spectral index; $h$ is the dimensionless Hubble parameter $h= H_0 / (100 \,{\rm km}\,{\rm s^{-1}}{\rm Mpc^{-1}})$; $M_{\rm \nu}$ is the mass of neutrinos in units of eV; and $w_0$ and $w_{\rm a}$ are parameters describing the time-evolving dark energy equation of state via $w(z) = w_0 + (1-a)\,w_{\rm a}$. We further consider a flat geometry and neglect the impact of radiation in the background evolution.

We note that this parameter volume coincides with that of the non-linear matter power spectrum and baryonic effects emulators of \cite{AnguloEtal2020} and \cite{AricoEtal2020}. We also recall that our suite of test simulations -- with which we validated the cosmology-rescaling approach, c.f. \S\ref{sec:validation} -- span the same range of parameters.

It is worth highlighting that although this represents a restricted parameter space, it is significantly larger than that typically covered by emulation of $N$-body simulations. This is because the cosmology-rescaling allows us to densely sample the volume, thus keeping emulation errors under control. Also, outside this parameter range, it could be possible to gracefully extrapolate using less precise methods. This could be acceptable as the region outside these boundaries is expected to be disfavored by large-scale-structure data, which would justify the use of less precise methods.

\subsection{Data}
\label{sec:data}

We sample our cosmology hyper-space with $400$ points generated with a latin-hypercube algorithm. For each point, we first select the simulation that would yield most accurate rescaling results \citep[see][for a details]{ContrerasEtal2020}, then apply our rescaling approach at $10$ different output times over $0<z<1.5$, and finally advect the corresponding Lagrangian fields. Note that the 10 redshifts considered are different for each cosmology, since they depend on the result of the minimization performed in the cosmology rescaling algorithm. The full sample, therefore, spans 4000 points in a 9-parameter hyperspace described by the 8 cosmology parameters presented in Eq. (\ref{eq:par_range}), plus the expansion factor (corresponding to redhsifts $0<z<1.5$).

For each rescaled output we measure the power spectra and compute the quantities to be emulated as we will describe in the next two subsections. This whole procedure takes approximately 2 CPU hours, with the largest fraction spent in the power spectra calculation and in the creation of the linear fields.

\subsubsection{Power Spectrum Measurements}

We compute 15 cross-spectra using a Fast Fourier Transform with $1080^3$ grid points and a Cloud-in-Cells mass assignment scheme. To reduce the impact of aliasing, we employ two interlaced grids, following \cite{SefusattiEtal2016}. We consider $50$ logarithmically-spaced bins over the range of wavenumbers $k \in [10^{-2}, 1]\,\ihMpc$.

On large scales, the regularity of the Fourier grid imprints features in the measured spectra. To correct for this, we compute the average of Fourier modes over the LPT expectation evaluated at the same wavelength, and then multiply by the LPT expectation at the bin centre, i.e.:

\begin{equation}
\hat{P}_{ij}(k) = P_{{\rm LPT}, ij}(k)
\sum_{|\vec{k}_n| \in [k-\Delta k, k + \Delta k]}
\frac{ \delta_i(\vec{k}_n) \delta_j(\vec{k}_n) }{P^{\rm LPT}_{ij}(\vec{k}_n)}
\end{equation}

\noindent we have found this procedure to be particularly important for the terms involving the $1$ and $\delta$ fields on large scales.

We perform the mass assignment and FFT using shared-memory parallelization over $24$ cores. The mass deposit steps typically take 8 seconds per field, and the corresponding FFT and power spectrum calculation approximately 17 seconds.

\subsubsection{Emulated quantities}

%============================
\begin{figure*}
\includegraphics[width=\textwidth]{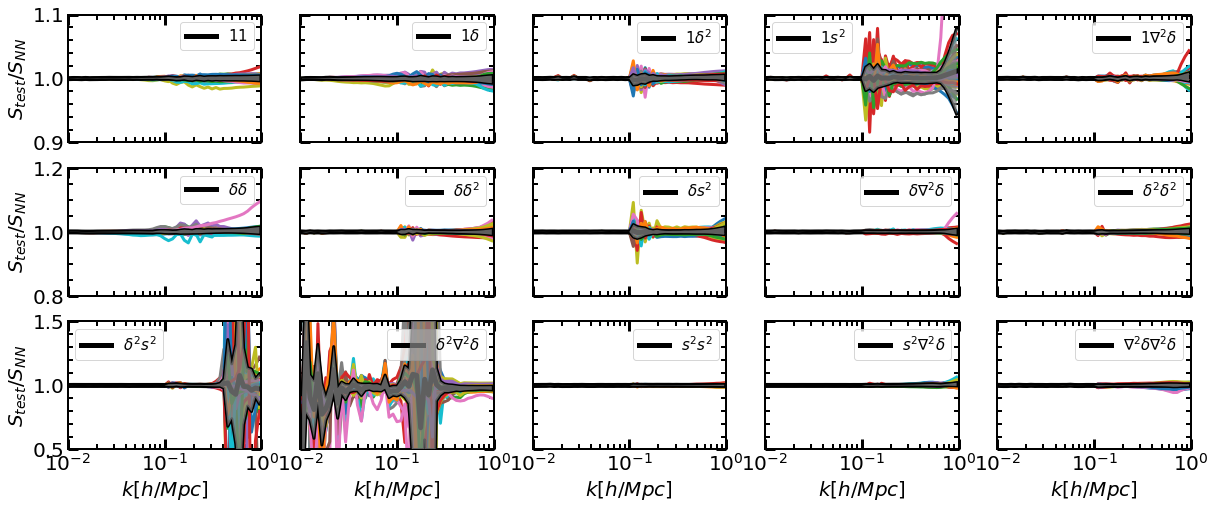}
\caption{ Ratio between the cross spectra in our validation set over the predictions of our neural network emulation. As in previous plots, each panel display results for a different combination of Lagrangian fields in Eulerian coordinates. Lines of different colour show specific test cosmologies, whereas shaded regions denote the are enclosing 95\% of the distribution with the mean indicated by the tick black line. Note the predictions are percent accurate over most cases, and the largest fractional discrepancies occur when the emulated quantities crosses zero. \label{fig:nn}}
\end{figure*}
%=============================

In total, we have computed a set of 15 spectra for $4000$ combinations of cosmology and redshifts.

In Fig.~\ref{fig:lpt} we show as an example one set of measurements for a randomly-selected cosmology at $z\sim0$. We also display our analytic LPT predictions as solid lines.

We can see that our measurements are in good agreement with LPT on large scales, as it was already shown in Fig.~\ref{fig:lpt-ratios}, especially for fields involving combinations of $1$, $\delta$, and $\delta^2$. In some spectra, there seems to be a small disagreement, which in \S\ref{sec:lpt} we attribute to either higher-order LPT contributions and/or shell-crossing. However, these later spectra seem to contribute subdominantly.

It is worth noting that there does not seem to be any significant noise in the fields that dominate the large-scale clustering $\langle 1 1\rangle$, $\langle 1 \delta \rangle$ and $\langle \delta \delta\rangle$. This owns to the \paf method, which, as shown by \cite{AnguloPontzen2016} \citep[see also][]{PontzenEtal2016,ChuangEtal2018,Villaescusa-NavarroEtal2018b}, results in <$\sim0.1\%$ accurate density power spectrum for $k<0.1\ihMpc$ at $z=0$ using simulations of comparable volume to those in the BACCO suite.

To reduce the dynamic range of our measurements, which would yield to a more accurate subsequent emulation, we consider the ratio of our measured cross-spectra over the corresponding analytic LPT expectation. We further consider the logarithm of the numerical and analytic results whenever that ratio is always positive over our range of wavenumbers, i.e:

\begin{equation}
S_{ij}(k,z) \equiv \begin{cases}
P_{ij} / P^{\rm LPT}_{ij}         ,& \text{if } \mathrm{min}[P_{ij}] \leq 0 \\
\log( P_{ij} / P^{\rm LPT}_{ij}) ,& \text{otherwise }
\end{cases}
\label{eq:sij}
\end{equation}

It is well known that large-scale coherent motions dilute the amplitude of the baryonic acoustic oscillations \citep[BAO][]{EisensteinEtal2007}. This effect can be captured in perturbation theory via the so-called infrared-resummation, however, this is not included in our LPT calculations. In principle, this is not a problem for our approach since we simply use LPT to reduce the dynamic range of our data. In practice, on the other hand, this means a small oscillatory feature is present in $S$ which we ought to emulate, which could introduce small uncertainties.

To avoid this and improve the accuracy of our predictions, in the definition of $\langle 1 1\rangle$, $\langle 1 \delta \rangle$ and $\langle \delta \delta\rangle$, we replace the corresponding LPT predictions by a linear theory power spectrum including an infra-red resummation, i.e:

\begin{equation}
P_{\rm linear}^{\rm smeared-BAO} \equiv P_{\rm linear}\,G(k) + P_{\rm linear}^{\rm no-BAO}\,[1 - G(k)]
\end{equation}

\noindent where $G(k) \equiv \exp[-k^2 / k_*]$, with $k_*^{-1} \equiv (6\pi^2)^{-1} \int {\rm d}k \,P_{\rm linear}$, and $P_{\rm linear}^{\rm no-BAO}$ is a version of the linear theory power spectrum without any BAO signal. Operationally, this is obtained by performing a discrete sine transform, smoothing the result, and returning to Fourier space by an inverse transform \citep{BaumannEtal2018,GiblinEtal2019}. We have checked that this in fact results into $S(k)$ devoid of any residual oscillations, however, we emphasise that our predictions are still essentially provided by our simulation results.

As discussed previously, simulations results have a significant amount of noise for some cross-spectra. Most notably, $1s^2$, $1\delta^2$, $\delta \delta^2$, and $\delta s^2$, in such cases emulation could be inaccurate as we would be trying to predict specific noise features. To avoid this, for the aforementioned spectra, we have replaced our measurements for $k<0.1\,\ihMpc$ with our analytic LPT spectra.

To further reduce the dimensionality and improve our emulation, we have performed a principal component decomposition over our dataset. We have found that keeping the first $6$ vectors with the largest eigenvalues is enough to explain most of the variance in all of our dataset.

\subsection{Neural Network}
\label{sec:emulation}

Following \cite{AnguloEtal2020} and \cite{AricoEtal2020}, we construct our emulators using a feed-forward Neural Network. We use a relatively simple architecture with two fully-connected hidden layers with 200 neurons each and a Rectified Linear Unit activation function. Note that a traditional alternative for building emulators are Gaussian Processes, which have the advantage of naturally providing with an estimate of the emulator uncertainty. These methods are, however, computationally expensive in high dimensions and for abundant training data. Thus, we choose to adopt neural networks instead to allow future expansions of the training set by including additional points in our parameter space, which could in principle reduce arbitrarily the emulation error.

We build a separate neural network for each cross-spectrum using the {\tt Keras} front-end of the {\tt Tensor-flow} library \citep{tensorflow2015-whitepaper}. We select a standard {\tt Adam} optimization algorithm with a $10^{-3}$ learning rate with a loss function given by the mean squared error.

We split our dataset, consisting of $4000$ sets of cross-spectra, into disjoint groups for training and validation. The training set comprises 95\% of the data with which the training of each 15 of the emulators takes approximately 30 minutes in a single Nvidia Quadro RTX 8000 GPU card; the evaluation of each emulator takes approximately $0.05$ seconds on the same hardware. The remainder of the full sample, 200 cross-spectra at cosmologies and redshifts never seen by the emulator during training, is used for validation.

\subsection{Validation}
\label{sec:validation}

We have estimated the accuracy of our neural network by comparing the prediction of our emulator with the measurements of the spectra in our test suite comprised of $\sim200$ different cosmologies and redshifts.

Our results are presented in Fig.~\ref{fig:nn}, where we display the ratio of the values of $S(k)$ predicted by our neural network and those directly measured in our test suite. Each panel shows the result for one of our 15 combinations of Lagrangian fields. We recall that in this testing set there are simulations with very dissimilar cosmologies including massive neutrinos and dark energy with time-dependent equation of state. Individual results are indicated by the coloured lines whereas the mean and the 95\% region of the distribution are indicated by thick solid and shaded regions.

We can see that our neural networks (NN) perform remarkably well overall. For the main fields for large-scale predictions -- i.e. combinations of $\delta$, $1$ -- the NN shows a typical uncertainty below 1\% at scales $k < 0.1\ihMpc$. On smaller scales, the accuracy somewhat degrades but it is still below 2\% for all but 1 case (pink line).

For fields further including $\delta^2$ terms, the accuracy is similar, with differences being less than 5\% on small scales. Recall that in the case of $1s^2$, $1\delta^2$, $\delta \delta^2$, and $\delta s^2$, we have replaced our numerical results on $k < 0.1\ihMpc$ by our analytic LPT expressions, which yields to an almost perfect emulation and to the apparent discontinuity of the emulator precision at that transition scale.

Note that the largest fractional errors occur at $k\sim0.2\ihMpc$ for $\delta^2 \nabla^2\delta$ and at $k\sim 0.7\ihMpc$ for $\delta^2 s^2$, which is where these cross-spectra cross zero (in many cosmologies), thus the ratio can easily become unbounded. However, since these fields are close to zero, their absolute contribution to the total power spectrum is negligible.

%improve further by considering more points

\section{Application: Fitting the galaxy power spectrum}
\label{sec:application}

%===========================
\begin{figure}
    \includegraphics[width=0.48\textwidth]{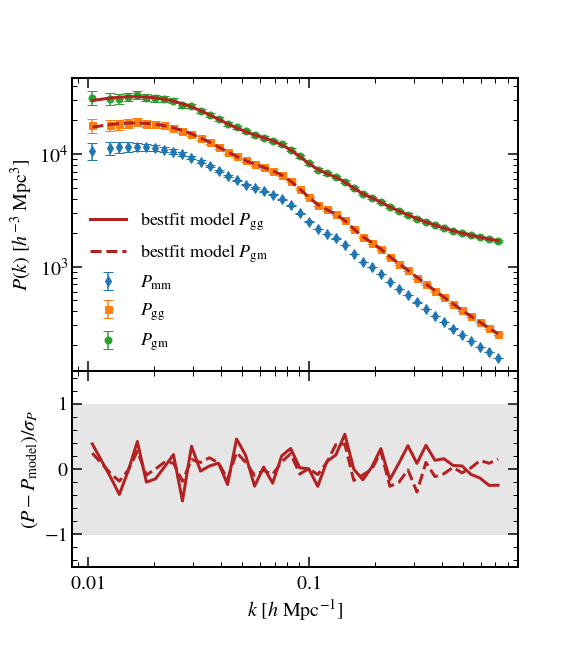}
    \caption{The galaxy auto power spectrum and matter-galaxy cross spectrum for SFR-selected mock galaxies with $\bar{n} = 10^{-3}\, h^3 \, \mathrm{Mpc}^{-3}$ at $z=1$. Symbols show the measurements for a galaxy-formation model in one of our BACCO simulations, whereas blue lines denote the best fitting model based on the emulator for biased tracers developed in our work. \label{fig::bestfit-model}}
\end{figure}
%============================

As an illustration of our emulator of the real-space clustering of biased tracers, in this section we will constrain the bias parameters that best fit the power spectrum of a simulated galaxy catalogue. While in principle we could directly jointly constrain bias parameters and cosmological parameters, for the sake of this paper we will fix our cosmology to the fiducial one. We do so to provide a quick test of the flexibility of the bias model that can be build with our emulator. In particular, by targetting possible inconsistencies between the inferred bias parameters as we include increasingly smaller scales in our analysis, we aim at showing that the hybrid lagrangian bias expansion model, as implemented in our emulator, is a good description of galaxy clustering down to unprecedentedly small scales. We defer the study of possible degeneracies between cosmology and Lagrangian bias parameters, and a thorough analysis of the limitations and performance of the model, to a future work.

%As an illustrative application of the method we infer the Lagrangian bias parameters that best fit a fiducial galaxy distribution as a function of the minimum scale included in the fit.

We consider a mock galaxy catalogue mimicking a sample of Emission Line galaxies at $z\sim1$ with a number density of $\bar{n} = 10^{-3}\, h^3 \, \mathrm{Mpc}^{-3}$, which can be regarded as analogous to the sample to be observed by {\it EUCLID} or {\it DESI}. To build this mock catalogue, we employ the extended subhalo abundance matching model of \cite{ContrerasAnguloZennaro2020b}. These authors showed that this particular implementation -- which includes models for tidal stripping and disruption of satellite galaxies, as well as for the star formation rate of galaxies -- accurately reproduces the real and reshift-space clustering of galaxies in the state-of-the-art hydrodynamical simulation TNG-300 \citep{NelsonEtal2018,SpringelEtal2018,MarinacciEtal2018,PillepichEtal2018,NaimanEtal2018}.

For this paper, we applied the aforementioned model to one of our main BACCO simulations ({\tt Nenya}), which is not a cosmology used in the training of the emulator. We employ the galaxy-formation parameters that best described SFR-selected TNG-300 galaxies, as provided by \cite{ContrerasAnguloZennaro2020b}. Specifically we set $\beta=2.490, \gamma=5.127, \log_{10} [M_1 / (h^{-1} \mathrm{M}_\odot)]=12.770, \tau_0=4.931,$ and $\tau_{\rm S}=-0.363$. We then selected the galaxies with the highest SFR values and kept a sample with a number density of $\bar{n} = 10^{-3}\, h^3 \, \mathrm{Mpc}^{-3}$. Please note that galaxies obtained with this method contain some galaxy assembly bias signal, since they inherit correlations with secondary halo properties (i.e. $v_{peak}$) and local environment \citep[see][for a more comprehensive treatment of GAB within SHAM methods]{ContrerasAnguloZennaro2020a}.

\begin{figure}
    \includegraphics[width=0.48\textwidth]{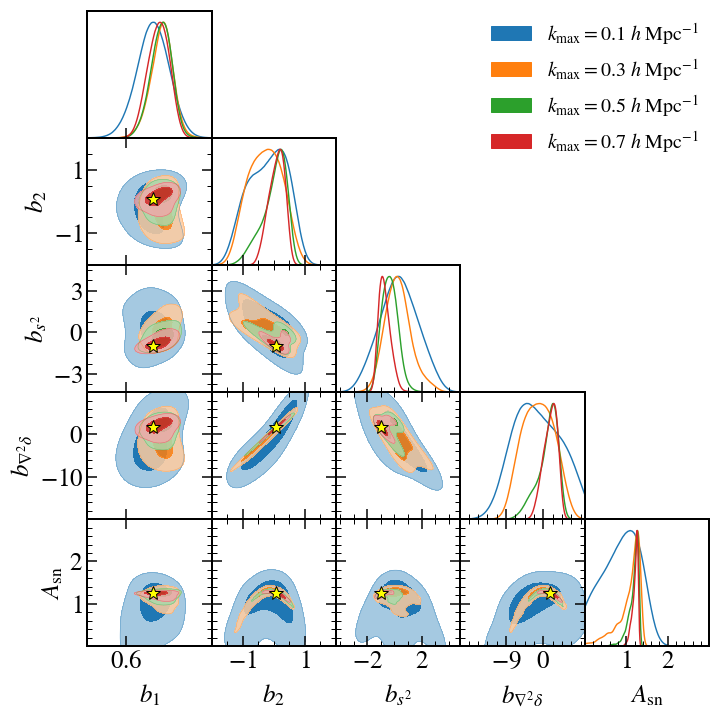}
    \caption{Marginalised 1$\sigma$ and $2\sigma$ credibility regions for the Lagrangian bias parameters $b_1$, $b_2$, $b_{s^2},$ and $b_{\nabla^2\delta}$ for a sample of mock galaxies with $\bar{n} = 10^{-3}\, h^3 \, \mathrm{Mpc}^{-3}$ selected according to their star formation rate at $z=1$. Yellow stars mark the MAP (Maximum A Posteriori) of the unmarginalised posterior. These constraints were derived by simultaneously fitting the galaxy power spectrum and galaxy-matter cross-spectrum from $k=10^{-2}\ihMpc$ until four different maximum wavenumbers, $k_{\rm max}$, $0.1$, $0.3$, $0.5$, and $0.7\ihMpc$, as indicated by the legend. \label{fig::contours}}
\end{figure}

We will consider as our data the average clustering measured in the two phase-inverted \nenya simulations to reduce the impact of cosmic variance on large scales. Specifically, we will employ the galaxy auto power spectrum, $P_{\rm gg}$, and the cross-correlation, $P_{\rm gm}$, with the underlying nonlinear matter field. As for the previous power spectrum calculations, we perform this operation using $1080^3$ FFTs with a CiC assignment and interlacing. In Fig.~\ref{fig::bestfit-model} we display the measurements as red symbols. Note that $P_{\rm gg}$ and $P_{\rm gm}$ involve different combinations of the Lagrangian fields, which makes it important to prove that the model is able to jointly fit them.

We will now explore the values of the free parameters of our real-space model at fixed cosmology, which is described by combinations of the cross spectra of Lagrangian fields and one scale independent shot-noise contribution:

\begin{eqnarray}
\label{eqs:model}
P_{\rm gg} =& \sum_{i,j \in \{1,\delta,\delta^2,\nabla^2\delta\}} b_i b_j\,P_{ij} +  \frac{A_{\rm sn}}{\bar{n}}  \\
P_{\rm gm} =& \sum_{i=1,j \in \{1,\delta,\delta^2,\nabla^2\delta\}} b_j\,P_{ij}
\end{eqnarray}

\noindent where $b_{i=1} \equiv 1$, $b_{i=\delta} \equiv b_1$, $b_{i=\delta^2} \equiv b_2$, $b_{i=s^2} \equiv b_{s^2}$, $b_{i=\nabla^2\delta} \equiv b_{\nabla^2\delta}$, and where $P_{ij}$ are the cross spectra predicted by our neural network for the cosmological parameters of the target galaxy catalogue (i.e. those of \nenya). Note that in principle, there could be other contributions to the noise \citep[e.g.][]{EggemeierEtal2020}, which could be included in our model but that we have neglected here. However, our model is flexible enough for it to be extended with additional terms. We defer an investigation of a minimal model for describing realistic galaxies to future work.  Our model is thus specified by 5 parameters:

\begin{equation}
\boldsymbol{\vartheta} = \{b_1, b_2, b_{s^2}, b_{\nabla^2\delta}, A_{\rm sn} \}
\end{equation}

\noindent where $b_i$ are the perturbative bias parameters of our sample. We assume flat priors for these parameters over the range $b_1 \in [-3, 3]$, $b_2 \in [-5,5]$, $b_{s^2} \in [-10,20]$, $b_{\nabla^2\delta} \in [-10,20]$, $A_{\rm sn} \in [0,2]$.

%To create the mock galaxy population we apply a SubHalo Abundance Matching (SHAM) to one of our main simulations at $z=0$, namely \nenya. In particular, following \cite{ContrerasAnguloZennaro2020b}, we use a set of SHAM parameters that reproduce the clustering of the TNG-300 simulation \matteo{cite}, with $\sigma_{M_*} = 0.1786, t_{\rm merger}=0.6617,$ and $f_s=0.0073$. After assigning a value of stellar mass to each subhalo, we extract a number density selected sample with $\bar{n}=0.01 \, h^3 \, \mathrm{Mpc}^{-3}$ \matteo{I'll change this for the SFR mock}.

%\begin{figure}
%    \includegraphics[width=0.48\textwidth]{figures/application/kmax_big_SFR_TNG_n_0.001_a_0.5.png}
%    \caption{The galaxy auto power spectrum and matter-galaxy cross spectrum for SFR-selected mock galaxies with $\bar{n} = 10^{-3}\, h^3 \, \mathrm{Mpc}^{-3}$ at $z=1$. Symbols show the measurements after a galaxy-formation model in one of our BACCO simulations, whereas blue lines denote the best fitting model based in the emulator for biased tracers developed in our work.
%    \label{fig::kmax}}
%\end{figure}

%We then proceed to jointly fit the galaxy-galaxy auto power spectrum and the galaxy-matter cross power %spectrum, assuming a diagonal covariance matrix, with errors corresponding to those expected for a {\it %Euclid}-like volume. To do so we compute\dots

%Using \texttt{emcee} \citep{Foreman-MackeyEtal2013} we maximize the posterior given by

We describe the probability of observing a set of power spectra $\tilde{P}(k) = \{P_{\rm gg}(k), P_{\rm gm}(k)\}$ as a multivariate normal distribution:

\begin{equation}
    \ln p[\tilde{P}, \boldsymbol{\vartheta}] \propto - \dfrac{1}{2} \boldsymbol{\Delta}^{\rm T} \, \mathcal{C}^{-1} \boldsymbol{\Delta} \, ,
\end{equation}
with
\begin{equation}
    \Delta(k_i) = \tilde{P}(k_i) - \tilde{P}_{\rm model}(k_i).
\end{equation}

Here we have assumed a block matrix covariance $\mathcal{C}$ \citep[each block corresponding to the auto-auto, cross-cross and auto-cross covariance, following][]{LiEtal2019} computed in the Gaussian limit with a volume corresponding to $V=25 h^{-3} \mathrm{Gpc}^3$, roughly comparable to the effective volume to be surveyed by EUCLID. To account for the uncertainty associated to our emulation process, we have added in quadrature a constant fractional error of $2\%$.

We sample the posterior probability of $\boldsymbol{\vartheta}$,

\begin{equation}
    p[\boldsymbol{\vartheta}, \tilde{P}(k)] = \dfrac{p[\tilde{P}(k), \boldsymbol{\vartheta}] p(\boldsymbol{\vartheta})}{p[\tilde{P}(k)]},
\end{equation}

\noindent using affine-invariant MonteCarlo Markov Chain code \texttt{emcee} \citep{Foreman-MackeyEtal2013}. At each step of the chain we evaluate our neural network emulator to obtain $S_{ij}$ (Eq. \ref{eq:sij}) for the given cosmology and redshift, and evaluate the corresponding LPT expressions to recover $P_{ij}$ and combine them as defined in Eqs.~\ref{eqs:model}. We consider $20$ MCMC walkers with 10,000 steps each and a burn-in phase of 500 steps. We checked that doubling the number of steps does not change significantly our results and therefore we consider our runs converged. Moreover, we have performed this analysis also employing the optimized simultaneous ellipsoidal nested sampler algorithm {\tt MultiNest} \citep{FerozEtal2009} with its python wrapper {\tt pymultinest} \citep{BuchnerEtal2014}. In this case we consider 400 live points with an evidence tolerance of 0.4 dex and we find results in agreement with our MCMC analysis.
%For each parameters, the Gelman-Rubin statistics was approximately, which suggest a good convergence of the chains.

%Here we have introduced the vector containing the auto- and cross-power spectrum $\tilde{P}(k) = \{P_{\rm gg}(k), P_{\rm gm}(k)\}$ and the vector of free parameters $\vartheta = \{b^{\rm L}_1, b^{\rm L}_2, b^{\rm L}_{s^2}, b^{\rm L}_{\nabla^2\delta}\}$.

In Fig. \ref{fig::contours} we present the two-dimensional marginalised constraints on the free parameters of our model for the galaxy power spectra. Dark and light regions denote the 1-$\sigma$ and 2-$\sigma$ two-dimensional confidence levels. We display the results for when limiting our analysis to different maximum wavenumbers, namely $k_{\rm max} = 0.1, 0.3, 0.5, 0.7 \hMpc$. We note that, as we include smaller scales, even though the width of the contours changes, the best fitting parameters remain compatible.

%To further reinforce this point, in Fig. \ref{fig::kmax} we show how the bestfitting parameters depend on our choice of $k_{\rm max}$. Until $k_{\rm max} = 0.75 \hMpc$, the inferred parameters so not show any scale dependence and we therefore consider all this range of scales suitable to obtain unbiased constraints. \matteo{compare with the large scale bias obtained from $P_{gg}/P_{gm}$}.

Finally, in Fig. \ref{fig::bestfit-model} we compare the measured galaxy statistics to those predicted by our model at the values that maximise the posterior probability.  We can see that our model is an excellent description of our simulated galaxy catalogues, over all the scales we consider and for both $P_{\rm gg}$ and $P_{\rm gm}$. In particular, our model evaluated for the best-fitting set of parameters is always within the error-bars of the data, and agrees with the measured power spectra at the $1-2\%$ level. This is particularly remarkable especially keeping in mind that, typically, perturbation-theory approaches are limited to scales $k \sim 0.2\ihMpc$, and that we compare against a galaxy catalogue built directly in Eulerian space and following physically-motivated recipes. We do not quote the value of our $\chi^2$ because our data vector is not a random sampling of our assumed covariance matrix, making the $\chi^2$ value itslef difficult to interpret. However, we refer the interested reader to \cite{ZennaroEtal2022}, where we present a thorough study of the model performance for different galaxy formation models, cosmologies, redshifts and number densities, showing that the obtained $\chi^2$ values always indicate the model to be a good fit for the data.

\section{Conclusions}
\label{sec:conclusions}

In this paper we have built and presented a set of 15 emulators for predicting the cross power spectra necessary to describe the galaxy-galaxy and galaxy-matter clustering in the context of a second order Lagrangian bias expansion. Each emulator predicts one such term on a range of wavenumbers $0.01 < k / \ihMpc < 1$ and at redshifts $0 < z < 1.5$, in cosmologies spanning $\sim 10 \sigma$ around the best-fitting parameters from the Planck collaboration, including extensions of the standard $\Lambda$CDM paradigm, such as massive neutrino and dynamical dark energy.

In order to build our emulator from a densely sampled parameter space, we have used a cosmology-rescaling algorithm. For this reason, we have first used a set of 35 intermediate volume simulations, with cosmologies spanning the same parameter space as our emulator, to show that the Lagrangian fields measured from simulations assuming different cosmologies can be accurately reproduced rescaling the small set of strategically chosen BACCO simulations presented in \cite{ContrerasEtal2020}. From this exercise (see Fig. \ref{fig:rescaling}) we conclude that the rescaling technique can accurately reproduce the Lagrangian fields needed to model Lagrangian bias expansion.

We have devoted particular care to obtaining theoretical predictions for the power spectra of these Lagrangian fields. We have considered LPT terms up to second order in density and advected them to Eulerian space. In Fig.\ref{fig:lpt-ratios} we have shown the accuracy of the expressions obtained this way, comparing them to numerical simulations of these LPT fields. We have used out LPT predictions for a two-fold purpose: one the one hand, we can combine the predictions to the actual data to mitigate the large-scale noise affecting some of the terms; on the other hand, we want to use them to reduce the dynamical range of the quantities that we emulate.

We have built our 15 emulators rescaling the large BACCO simulations to approximately 4000 different combinations of cosmology and redshifts (see Fig. \ref{fig:lpt}) and we have then proceeded to validate them against a set of 200 combinations of cosmologies and redshifts (Fig. \ref{fig:nn}). We find that most of the terms that principally contribute to the galaxy power spectrum (namely, the terms involving combinations of the homogeneous and linear fields, $1, \delta$) are emulated with percent accuracy. The terms involving the squared density field $\delta^2$, the tidal field $s^2$ and the laplacian of the density field $\nabla^2\delta$ are generally reproduced with an accuracy of a few percent. However, two of these terms, the combination of $\delta^2 s^2$ and of $\delta^2 \nabla^2\delta$, show particularly bad performances at the scale where they cross zero.

Finally, we have used our emulator to constrain the bias parameters of a mock galaxy sample of known cosmology, using the galaxy-galaxy and galaxy-matter power spectra. We constructed our galaxy sample at redshift $z=1$, selecting galaxies according to their Star Formation Rate until reaching a number density of $0.001 h^3 \, \mathrm{Mpc}^{-3}$ and assigning errors that reflect the cosmic variance expected for large upcoming surveys such as Euclid. This proof of concept showed that this technique can be used to model the clustering of realistic galaxy samples, obtaining constraints of the bias parameters that are self consistent along a large range of scales (Fig. \ref{fig::bestfit-model}). In particular we were able to fit our mock galaxy sample including scales down to $k_{\rm max}=0.7\ihMpc$.

We anticipate this approach will be valuable for studying Lagrangian biases from realistic galaxy mock samples. Its applications could be extended also to any problem in which the knowledge of the fully nonlinear clustering of each of these lagrangian fields is required, as for example in the modelling of galaxy-galaxy lensing. However non trivial, we also expect that an extension of this emulator to redshift space would provide us with an unprecedented tool to constrain cosmology from galaxy clustering.

In the latest stages of the preparation of this work, a similar approach to ours has been made public in \cite{KokronEtal2021}. The two works share many similarities, although differ in many aspects of the implementation and analysis. We defer to future works a comparison of the two approaches.

\section*{Acknowledgments}

The authors acknowledge the support of the ERC-StG number 716151 (BACCO). MPI acknowledges the support of the ``Juan de la Cierva Formaci\'on'' fellowship (FJC2019-040814-I). SC acknowledges the support of the ``Juan de la Cierva Incorporaci\'on'' fellowship (IJC2020-045705-I). The authors acknowledge the computer resources at MareNostrum and the technical support provided by Barcelona Supercomputing Center (RES-AECT-2019-2-0012, RES-AECT-2020-3-0014).

\section*{Data Availability}

The data underlying this article will be shared on reasonable request to the corresponding author. The Neural Network emulator will be made public at \url{http://www.dipc.org/bacco} upon the publication of this article.



\bibliographystyle{mnras}
\bibliography{Bibliography_all} % if your bibtex file is called example.bib

%%%%%%%%%%%%%%%%%%%%%%%%%%%%%%%%%%%%%%%%%%%%%%%%%%

%%%%%%%%%%%%%%%%% APPENDICES %%%%%%%%%%%%%%%%%%%%%

\appendix

\section{LPT power spectra} \label{sec::lpt-predictions}
In Eq. (\ref{eq::pt-terms-general}) we presented a general form
expressing the cross power spectra $P_{ij}$ of the different fields, with $i,j = \{ 1, \delta, \delta^2, s^2, \nabla^2\delta \}$. Here, we report the prefactors of the terms of order $(11)$ and $(22)$, and the different kernels entering the $(22)$ terms.

The $(11)$ terms are multiplied by these prefactors (note the $k$-dependence arising for terms including the Laplacian of the density field),

\begin{equation}
    A_{ij}(k) = \begin{bmatrix}
        1    & 1    & 0 & 0 & -k^2 \\
        1    & 1    & 0 & 0 & -k^2 \\
        0    & 0    & 0 & 0 & 0    \\
        0    & 0    & 0 & 0 & 0    \\
        -k^2 & -k^2 & 0 & 0 & k^4  \\
    \end{bmatrix}.
\end{equation}

The $(22)$ terms are preceded by
\begin{equation}
    B_{ij}(k) = \begin{bmatrix}
        1 & 1   & 1 & 1 & 0 \\
        1 & 1/2 & 2 & 2 & 0 \\
        1 & 2   & 1 & 1 & 1 \\
        1 & 2   & 1 & 1 & 1 \\
        0 & 0   & 1 & 1 & 1 \\
    \end{bmatrix}.
\end{equation}

% \begin{widetext}
% \begin{equation}
%     C_{ij}(\pp, \kk)\begin{bmatrix}
%         F_{\rm ZA}^2(\qq, \kk-\qq) & F_{\rm ZA}(\qq, \kk-\qq) \left[\dfrac{\kk \cdot (\kk-\qq)}{|\kk-\qq|^2} + \dfrac{\kk \cdot \qq}{q^2}\right] & F_{\rm ZA}(\qq, \kk-\qq) & F_{\rm ZA}(\qq, \kk-\qq) S_2(\qq, \kk-\qq) & 0 \\
%          & \dfrac{\kk \cdot (\kk-\qq)}{|\kk-\qq|^2} \left[\dfrac{\kk \cdot (\kk-\qq)}{|\kk-\qq|^2} + \dfrac{\kk \cdot \qq}{q^2}\right] & \dfrac{\kk \cdot (\kk-\qq)}{|\kk-\qq|^2} & \dfrac{\kk \cdot (\kk-\qq)}{|\kk-\qq|^2} S_2(\qq, \kk-\qq) & 0 \\
%          & & 1 & S_2(\qq, \kk-\qq) & \left[q^2 \dfrac{\kk \cdot (\kk-\qq)}{|\kk-\qq|^2} + |\kk-\qq|^2 \dfrac{\kk \cdot \qq}{q^2}\right] \\
%          & & & \left[q^2 \dfrac{\kk \cdot (\kk-\qq)}{|\kk-\qq|^2} + |\kk-\qq|^2 \dfrac{\kk \cdot \qq}{q^2}\right] S_2(\qq, \kk-\qq)\\
%          & & & & \left[\dfrac{q^4}{|\kk-\qq|^4}(k^2 - \kk\cdot\qq)^2 + \kk \cdot \qq (k^2 - \kk \cdot \qq)\right] \\
%     \end{bmatrix}
% \end{equation}
% \end{widetext}

Finally, we report all the kernels for the $(22)$ terms. Please note that we use

\begin{equation}
    F_{\rm ZA}(\kk_1, \kk_2) = 1 + \dfrac{\kk_1 \cdot \kk_2}{k_1 k_2} \left(\dfrac{k_1}{k_2} + \dfrac{k_2}{k_1}\right) + \left(\dfrac{\kk_1 \cdot \kk_2}{k_1 k_2}\right)^2,
\end{equation}

\noindent and

\begin{equation}
    S_2(\kk_1, \kk_2) = \left(\dfrac{\kk_1 \cdot \kk_2}{k_1 k_2}\right)^2 - \dfrac{1}{3}.
\end{equation}

\begin{equation}
    C_{11}(\pp, \kk) = F_{\rm ZA}^2(\pp, \kk-\pp)
\end{equation}
\begin{equation}
    C_{1\delta}(\pp, \kk) = F_{\rm ZA}(\pp, \kk-\pp) \left[\dfrac{\kk \cdot (\kk-\pp)}{|\kk-\pp|^2} + \dfrac{\kk \cdot \pp}{p^2}\right]
\end{equation}
\begin{equation}
    C_{1\delta^2}(\pp, \kk) = F_{\rm ZA}(\pp, \kk-\pp)
\end{equation}
\begin{equation}
    C_{1 s^2}(\pp, \kk) = F_{\rm ZA}(\pp, \kk-\pp) S_2(\pp, \kk-\pp)
\end{equation}
\begin{equation}
    C_{1 \nabla^2\delta}(\pp, \kk) = 0
\end{equation}
\begin{equation}
    C_{\delta\delta}(\pp, \kk) = \dfrac{\kk \cdot (\kk-\pp)}{|\kk-\pp|^2} \left[\dfrac{\kk \cdot (\kk-\pp)}{|\kk-\pp|^2} + \dfrac{\kk \cdot \pp}{p^2}\right]
\end{equation}
\begin{equation}
    C_{\delta\delta^2}(\pp, \kk) = \dfrac{\kk \cdot (\kk-\pp)}{|\kk-\pp|^2}
\end{equation}
\begin{equation}
    C_{\delta s^2}(\pp, \kk) = \dfrac{\kk \cdot (\kk-\pp)}{|\kk-\pp|^2} S_2(\pp, \kk-\pp)
\end{equation}
\begin{equation}
    C_{\delta \nabla^2\delta}(\pp, \kk) = 0
\end{equation}
\begin{equation}
    C_{\delta^2\delta^2}(\pp, \kk) = 1
\end{equation}
\begin{equation}
    C_{\delta^2 s^2}(\pp, \kk) = S_2(\pp, \kk-\pp)
\end{equation}
\begin{equation}
    C_{\delta^2 \nabla^2\delta}(\pp, \kk) = \left[p^2 \dfrac{\kk \cdot (\kk-\pp)}{|\kk-\pp|^2} + |\kk-\pp|^2 \dfrac{\kk \cdot \pp}{p^2}\right]
\end{equation}
\begin{equation}
    C_{s^2 s^2}(\pp, \kk) = S_2(\pp, \kk-\pp)^2
\end{equation}
\begin{equation}
    C_{s^2 \nabla^2\delta}(\pp, \kk) = \left[p^2 \dfrac{\kk \cdot (\kk-\pp)}{|\kk-\pp|^2} + |\kk-\pp|^2 \dfrac{\kk \cdot \pp}{p^2}\right] S_2(\pp, \kk-\pp)
\end{equation}
\begin{equation}
    C_{\nabla^2\delta \nabla^2\delta}(\pp, \kk) = \left[\dfrac{p^4}{|\kk-\pp|^4}(k^2 - \kk\cdot\pp)^2 + \kk \cdot \pp (k^2 - \kk \cdot \pp)\right]
\end{equation}

\section{Comparison against public PT codes} \label{sec::velocileptors-fastpt}

In this appendix we compare our implementation of LPT integrals and the resulting prediction for Lagrangian fields in Eulerian space against those computed using two publicly available codes.

Specifically, we compare against {\tt FastPT}, which implements standard perturbation theory at 1 loop, and against {\tt velocileptors}, which employs 1-loop Lagrangian perturbation theory with effective-field theory counterterms. Note that not all power spectrum combinations are available in those codes, thus we restrict the comparison to a subset of integrals.

\begin{figure*}
    \includegraphics[width=\textwidth]{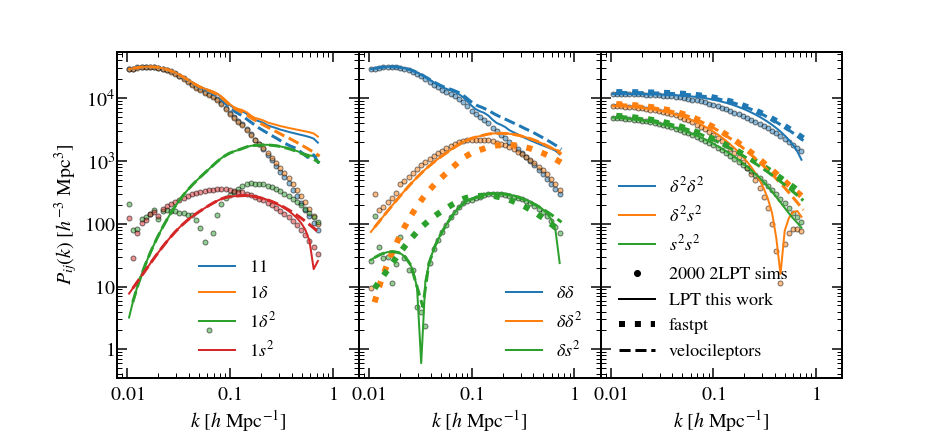}
    \caption{Various cross-spectra, $P_{ij}$ at $z=0$, as indicated by the legend. Shaded areas represent the standard deviation of 2000 fast simulations evolved with 2LPT. Solid lines show our results, whereas dashed lines and crosses denote the values obtained using {\tt velocileptors} and {\tt FastPT}, respectively. \label{fig::fastpt-velocileptor}}
\end{figure*}

We first compare our LPT solutions against {\tt velocileptors}. We see that, on large scales, there is a good agreement for $11$, $1 \delta$ $\delta \delta$, $\delta \delta^2$, $\delta s^2$, $\delta^2 \delta^2$, $\delta^2 s^2$, and $s^2 s^2$. We find some differences in the large scale behavior of the terms $1 \delta^2$ and $1 s^2$ that we plan to ivestigate in the future. When compared against {\tt FastPT}, we also see a good agreement with our predictions on large scales for $\delta \delta$, $\delta^2 \delta^2$ and $s^2 s^2$, but systematic differences for $\delta \delta^2$, $\delta s^2$.

Our ensemble of LPT simulations shows very good agreement with our predictions (as well as with those measured in full $N$-body simulations (c.f. \S\ref{sec:lpt}). Note that, for the term $1 \delta^2$, the LPT simulations appear to have less power than the predictions. However, for this term, the 2LPT simulations also exhibits less power than the $N$-body solution, that in turn is in better agreement with our predictions, as shown in Fig. \ref{fig:lpt-ratios}.

%%%%%%%%%%%%%%%%%%%%%%%%%%%%%%%%%%%%%%%%%%%%%%%%%%

% Don't change these lines
\bsp	% typesetting comment
\label{lastpage}
\end{document}